\documentclass[prl,aps,showpacs,twocolumn,superscriptaddress]{revtex4-1} \input epsf
\usepackage{graphicx}
\usepackage{color}
\usepackage{dcolumn}
\usepackage{bm}
\usepackage{amsmath}
\usepackage[normalem]{ulem}

\begin{document}
\preprint{APS/123-QED}
\title{Minimal Lattice Model of Lipid Membranes with Liquid-Ordered Domains}
\author{Tanmoy Sarkar}
\affiliation{Department of Biomedical Engineering, Ben Gurion University of the Negev, Be'er Sheva 84105, Israel}
\author{Oded Farago}
\email{ofarago@bgu.ac.il}
\affiliation{Department of Biomedical Engineering, Ben Gurion University of the Negev, Be'er Sheva 84105, Israel}
\date{\today}
\begin{abstract}

Mixtures of lipids and cholesterol are commonly used as model systems
for studying the formation of liquid-ordered ($L_o$) domains in
heterogeneous biological membranes. The simplest model system
exhibiting coexistence between $L_o$ domains and a liquid-disordered
($L_d$) matrix is that of a binary mixture of saturated lipids like
DPPC and cholesterol (Chol). DPPC/Chol mixtures have been investigated
for decades both experimentally, theoretically, and recently also by
means of atomistic simulations. Here, we present a minimal lattice
model that captures the correct behavior of this mixture across
multiple scales. On the macroscopic scales, we present simulation
results of mixtures of thousands of lipids and Chol molecules which
show excellent agreement with the phase diagram of the system. The
simulations are conducted on timescales of hundreds of microseconds
and show the morphologies and dynamics of the domains. On the
molecular scales, the simulations reveal local structures similar to
those recently seen in atomistic simulations, including the formation
of gel-like nano-domains ($\sim$ 1-10 nm) within larger Chol-rich
$L_o$ domains ($\sim$ 10-100 nm).  The observed multi-scale behavior
is related to the tendency of Chol to induce ordering of acyl chains
on the one hand, and disrupt their packing with each other, on the
other hand.
\end{abstract}


\maketitle


{\bf Introduction:} The plasma cell membrane is characterized by a
lateral heterogeneous distribution of its biomolecular
components~\cite{alberts-book}. Specifically, it contains small
(10-200 nm) domains, termed lipid rafts, enriched in saturated lipids
(mainly sphingolipids), cholesterol (Chol), and often particular
proteins~\cite{simons97,pike06}. While many aspects of this ``raft
hypothesis'' (including even their very existence in cellular
membranes) remain controversial~\cite{levental20}, there are strong
evidences that raft domains play a key role in cellular processes such
as signal transduction~\cite{simons00}, cell
adhesion~\cite{leitinger02,murai12}, and membrane
trafficking~\cite{ikonen01,hanzal07}. Rafts are ``liquid-ordered''
($L_o$) domains where the lipid hydrocarbon chains are more ordered
and more tightly packed than the surrounding non-raft
``liquid-disordered'' ($L_d$) phase of the
bilayer~\cite{vdgoot01,kaiser09}.  The $L_o$ domains have features
resembling both the $L_d$ phase (lateral lipid
mobility~\cite{filippov04}) and the ``gel'' phase where the lipid acyl
chains are fully extended, tightly packed, and spatially
ordered~\cite{holl08}.

\begin{figure*}
\centering
\includegraphics[width=16cm]{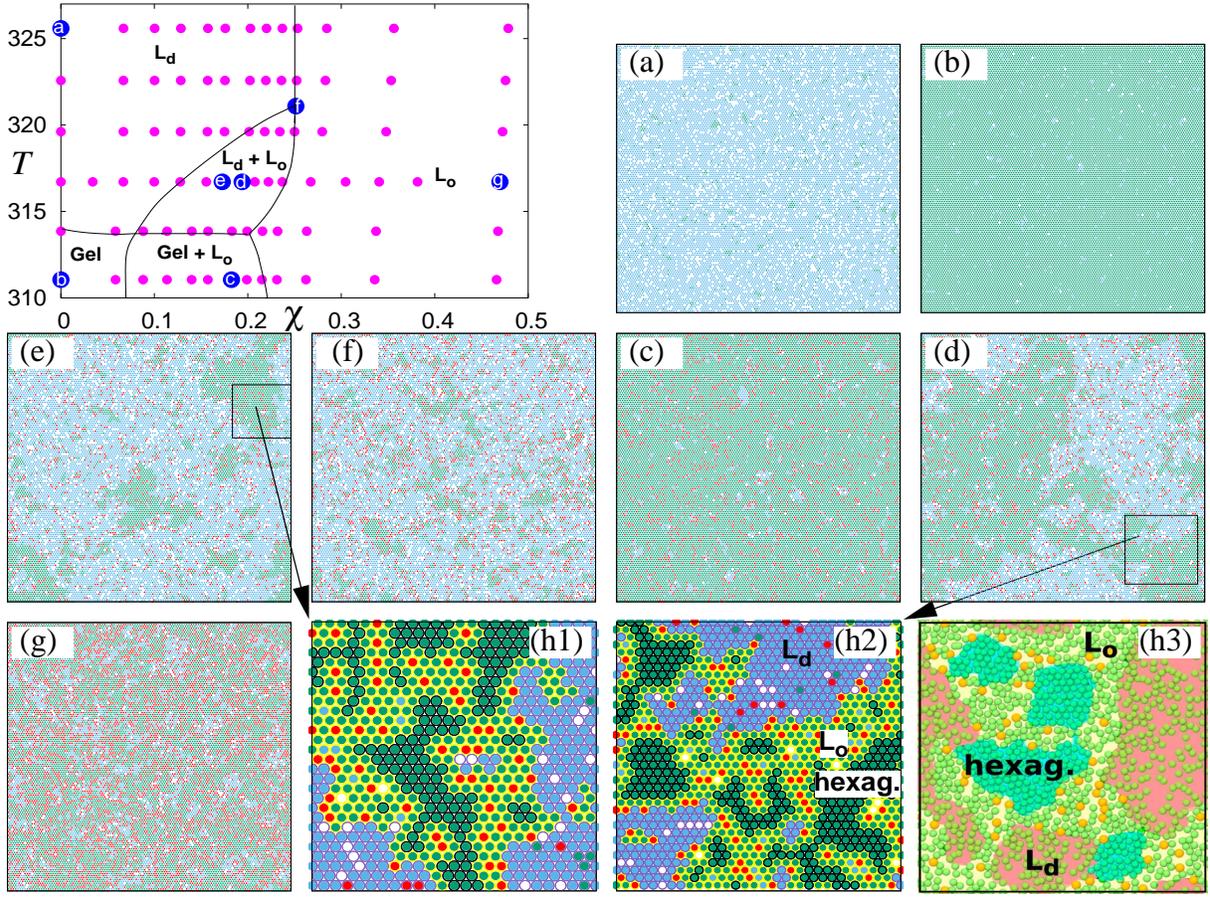}
\caption{Top left: The phase diagram of DPPC/Chol mixtures, with phase
  boundaries (solid black lines) taken from ref.~\cite{vist90}.
  Magenta dots show locations of the simulated systems.  The blue dots
  with white letters denote the locations in phase space of the
  systems from which snapshots (a)-(g) are taken. In the snapshots,
  ordered and disordered lipid chains, cholesterol, and empty sites
  (voids) are colored in green, blue, red and white,
  respectively. (h1,h2) Zoom-in-views of certain regions in snapshots
  from the $L_d$+$L_o$ region [black boxes in (e),(d)]. Here, the
  sites belonging to the gel, $L_o$, and $L_d$ are marked with black,
  yellow, and purple borders, respectively. (h3) A snapshot taken
  from atomistic simulations presented in
  ref.~\cite{javanainen17}. Here, DPPC chains and cholesterol are
  shown in green and orange, respectively, and the hexagonal (hexag.),
  $L_o$, and $L_d$ regions are, respectively, shaded with blue,
  yellow, and red.}
\label{fig1}
\end{figure*}

Biophysical studies of $L_o$ domains are rarely performed on
biological membranes that contain hundreds types of lipids. Instead,
simple model systems have been investigated, typically ternary
mixtures of saturated and unsaturated lipids and
cholesterol~\cite{komura04,feigenson07,veatch07,goni08,putzel08,chong09,feigenson09,schick12,komura14,sadeghi14,sodt14}.
An even simpler system exhibiting domain formation is a binary mixture
of Chol and saturated lipids like DPPC. Specifically, $L_o$ domains
appear in coexistence with an $L_d$ phase at intermediate Chol
concentrations and temperatures slightly above the DPPC liquid-gel
transition temperature, $T_m$~\cite{ispen87, vist90} (see phase
diagram in fig.~\ref{fig1}).  The domains appear to be both small and
transient, not macroscopically phase separated from the $L_d$
matrix~\cite{schmid17}. Recently, the homogeneity of the $L_o$ phase
itself has been called into question~\cite{rhein13}.  Neutron
scattering data at high Chol concentration has been interpreted as if
the $L_o$ phase contains highly ordered nano-domains residing in a
somewhat less ordered environment~\cite{amstrong13}.  A recent
atomistic simulation study of DPPC/Chol mixtures has provided further
support to this picture~\cite{javanainen17}. The simulations reveal
the existence of Chol-free hexagonally-packed small clusters of acyl
chains within liquid ordered domains that are rich in Chol, especially
along the domain boundaries. These new findings revoke a renewed
discussion on the phase behavior of DPPC/Chol mixtures.

In this letter, we present a lattice model addressing the multi-scale
behavior of DPPC/Chol mixtures.  It reproduces the local inhomogeneous
structure of the domains on the one hand, and the phase diagram of the
whole system on the other hand. The model features a considerably
smaller number of parameters compared to classical lattice models of
DPPC/Chol mixtures like the 10-state Potts-like Pink
model~\cite{pink80} and the model of Ipsen et al.~\cite{ispen87}. It
is also different than Almeida lattice model~\cite{almeida11} that
uses a top-down thermodynamic approach that assigns free energies to
each state (gel/liquid disordered/liquid ordered). A unique feature of
the~model presented herein is the presence of empty sites that
represent small area voids. This facilitates investigations of the
diffusive dynamics of the lipids in the different phases. The minimal
nature of the model and its ability to provide multi-scale information
on both structural and dynamic properties of the system helps us to
unravel the main thermodynamic and molecular forces underlying the
formation of liquid order domains. Specifically, we identify the
exclusion of Chol from both the gel and the $L_d$ phases as the reason
for its accumulation along the interfaces between them. The
accumulated high concentration of Chol molecules triggers the ordering
of the lipid chains in their vicinity.

{\bf Model and Simulations:} DPPC lipids are modeled as dimers of two
chains, while the Chol molecules are represented by single
chains. Each site of the lattice can be in one of four states
($s=0,1,2,3$): It can be occupied by a DPPC chain in either a
disordered ($s=1$) or an ordered ($s=2$) state, a cholesterol molecule
($s=3$), or be empty ($s=0$) representing a small area void.  The
introduction of empty sites (i) allows the Monte Carlo (MC)
simulations to mimic the diffusive dynamics of the lipids, and (ii)
enables setting correctly the area per lipid in the different states
(see details below), rather than making it a state-dependent
adjustable parameter (as in~\cite{ispen87,pink80}). This is also the
reason for modeling lipids as dimers rather than monomers.  The
disordered state of the DPPC chain ($s=1$) is entropically favored by
a free energy $F_1=-\Omega_1 k_BT$ (where $k_B$ is Boltzmann constant
and $T$ is the temperature) compared to the ordered state ($s=2$), but
two nearest neighbor chains in the ordered state have interaction free
energy $F_{22}=-\epsilon_{22}<0$ representing the favorable packing of
chains in this state. The competition between these terms drives a
first order melting transition of the DPPC bilayer at
$T_m=314K$~\cite{klump81}. Chol interaction free energy with a nearest
neighbor ordered chain is equal to $F_{23}=-\epsilon_{23}<0$.  We set
$\epsilon_{23}<\epsilon_{22}$, which means that Chol prefers locations
close to the ordered chains, but with interaction free energy that is
weaker than their packing interactions with each other. The physical
length of the lattice spacing, $l$, is determined by considering a
pure DPPC system (without Chol). In the gel state, we expect the
ordered lipid chains to cover almost the entire lattice with only a
negligible fraction of empty lattice sites. Thus, the lattice spacing
is set to $l=5.5~{\AA}$, which means that the area per lipid (which
occupies two sites) is $a\simeq 52~{\AA}^2$~\cite{javanainen17}, and
the area per Chol is $a_C\simeq 26~{\AA}^2$~\cite{smond99}.  In the
$L_d$ state, the area per DPPC molecule increases to roughly $a\simeq
60~{\AA}^2$~\cite{javanainen17}. This means that about $15\%$ of the
lattice sites in the $L_d$ phase must be empty.  In order to
accomplish this, we introduce an elasticity-like quadratic term with a
minimum at a ratio of $c_r=(85/2)/15\simeq 2.83$ between the number of
lipids with two disordered chains ($N_{11}$) and the number of voids
($N_0$). Taking all together, the energy assigned to each MC
configuration is given by
\begin{eqnarray}
E&=&-\Omega_1 k_BT\sum_{i}
\delta_{s_i,1}-\sum_{i,j}\epsilon_{22}\delta_{s_i,2}\delta_{s_j,2}
\label{eq:mcenergy}\\ 
&-&\sum_{i,j} \epsilon_{23}\left[\delta_{s_i,2}\delta_{s_j,3}
  +\delta_{s_i,3}\delta_{s_j,2}\right] +e_m(N_{11}-c_rN_0)^2/N_s,\nonumber
\end{eqnarray}
where the deltas are Kronecker deltas, and the summations are carried
over all $N_s$ lattice sites in the first term and over all nearest
neighbor pair sites in the second and third terms in
Eq.~(\ref{eq:mcenergy}).  Defining an energy unit $\epsilon=1$, we set
the packing energies in Eq.~(\ref{eq:mcenergy}) to
$\epsilon_{22}=1.3\epsilon$ and $\epsilon_{23}= 0.72\epsilon$. The
entropy parameter in the first term in Eq.~(\ref{eq:mcenergy}) is set
to $\Omega_1=3.9$, and the elastic modulus in the fourth term is
$e_m=85\epsilon$. These values are in reasonable consistency with
values of related parameters used in classical models that inspired
our work~\cite{ispen87,doniach78,almeida09,almeida2018}.

The MC simulations are conducted on a triangular lattice of
$N_s=121\times 140=16940$ sites (with periodic boundary conditions)
that has an aspect ratio close to~1. They involve three types of move
attempts - (i) displacements of randomly chosen Chol or a lipid chain
(dimer rotation) to a nearest neighbor vacant site, (ii) exchange of a
lipid having two ordered chains with a pair of nearest neighbor voids
and vise versa ($22\leftrightarrow00$: density-changing), and (iii)
flipping the state (ordered/disordered) of a randomly chosen lipid
chain ($1\leftrightarrow 2$: state-exchange).  In a pure DPPC systems,
we start by dividing the system into two halves, one in the gel and
one in the liquid-disordered phase. The system is equilibrated until
one of these phases takes over the other. As expected, the melting
transition is first order and takes place at
$k_BT_m\simeq0.90\epsilon$. Chol is added by taking an equilibrated
pure system and replacing randomly chosen lipids with two Chol chains,
after which the system is further simulated until it settles into the
new equilibrium distribution.

{\bf Results:} We simulated the system at different temperatures ($T$)
and different molar fractions of Chol ($\chi$). The phase diagram is
shown in the upper left corner of fig.~\ref{fig1}, where the magenta
dots indicate the locations in phase space of the simulated systems,
and the blue dots the locations of systems from which snapshots
(a)-(g) are taken. The phase boundaries (solid black lines) are taken
from ref.~\cite{vist90}, and we find our observations to be consistent
with them. At low concentrations of Chol, the system is in the gel
phase for $T<T_m$ [fig.~\ref{fig1}(b)] and in the $L_d$ phase for
$T>T_m$ [\ref{fig1}(a)]. In the former the lattice is almost fully
covered by ordered chains (depicted in green), while in the latter the
lattice is mostly covered by disordered chains (blue) with a visible
fraction of empty sites (white). At high Chol (red) concentrations
($\chi\gtrsim 0.25$), the system is found in a single $L_o$ state,
containing a relatively high fraction of ordered chains
[\ref{fig1}(g)]. Snapshots (d), (e) and (f), which are from the
$L_d+L_o$ coexistence region, clearly show the appearance of domains
of ordered lipids within a sea of disordered chains.  These domains
are dynamic and have irregular morphologies, suggesting that the
system is {\em not}\/ macroscopically phase separated.

\begin{figure*}
\centering
\includegraphics[width=18cm]{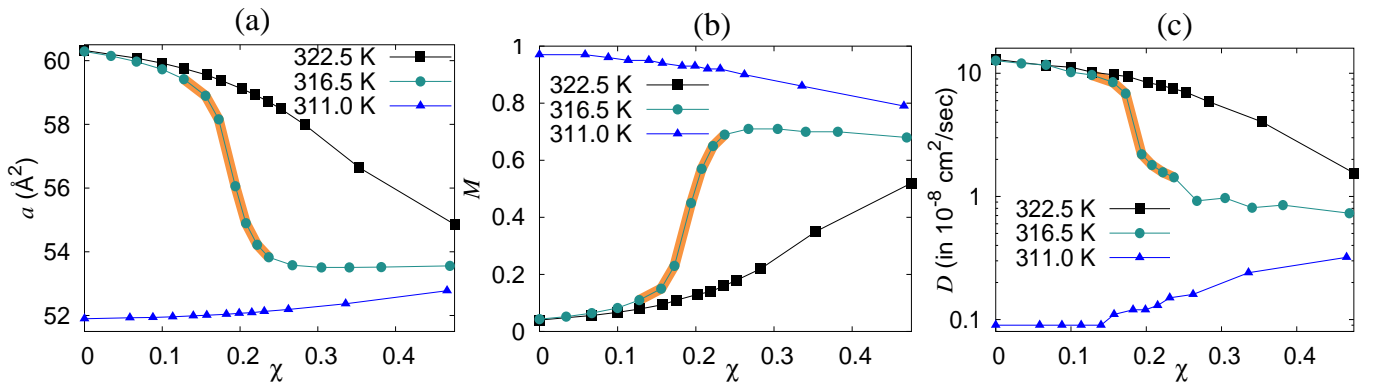}
\caption{(a) The area per lipid, (b) fraction of chains in the ordered
  state (the ratio between the number of ordered chains and the total
  number of lipid chains), and (c) average diffusion coefficient of
  the lipids, as a function of the cholesterol mole fraction.
  At $T=316.5{\rm K}$, the system enters the
  $L_d$+$L_o$ two phase region at intermediate cholesterol
  concentrations, which is indicated by the portion of the curves
  shaded with orange.}
\label{fig2}
\end{figure*}

In order to quantify the properties of the different phases, we
measured the area per lipid, $a=52~{\AA}^2\times$(\# of lipid chains
and voids/\# of lipid chains) [fig.~\ref{fig2}(a)], and the fraction
of ordered chains, $M$ [\ref{fig2}(b)], at three different
temperatures. For $T=311K<T_m$ (blue triangles), we observe a gradual
increase in $a$ and a gradual decrease in $M$ with $\chi$. For
temperature $T=322.5K$ (black squares), which is substantially above
$T_m$, we see the opposite trends, namely a decrease in $a$
accompanied by an increase in $M$. These trends reflect the fact that
the level of order and chain packing in the liquid ordered state is
intermediate between the gel and liquid disordered states. For
$T=316.5K\gtrsim T_m$ (green circles), a sharp crossover from $L_d$
(high $a$, low $M$) to $L_o$ (Low $a$, high $M$) is observed for
$0.12\lesssim\chi\lesssim 0.24$, which is exactly the location of the
$L_d+L_o$ coexistence regime in phase space.  Similar trends are also
detected in fig.~\ref{fig2}(c) which shows the average diffusion
coefficient, $D$, of the lipids. In full agreement with atomistic
simulation results~\cite{javanainen17}, we find the value of $D$ in
the liquid disordered state to be two orders of magnitude larger than
that of the gel where the mobility of the lipids is nearly
non-existing. In the liquid ordered state, $D$ is about an order of
magnitude smaller (larger) than in the liquid disordered (gel) state,
implying that this is a relatively viscous liquid.

The values of $D$ in fig.~\ref{fig2}(c) are quoted in physical time
units that are related to the MC time by using the experimental value
of $D=11.2\times 10^{-8} cm^2/s$ at $T=322.5K$ and $\chi=0.1$ ($L_d$
state)~\cite{javanainen17}. The simulations extend over $\sim$ 5 mses,
i.e., 3-4 orders longer than the atomistic simulations
in~\cite{javanainen17}. The Supplemental Material movie~\cite{supp}
shows $100\ \mu s$ dynamics (with a time difference of $4.6\ \mu s$
between consecutive frames) of liquid ordered domains in the $L_d+L_o$
coexistence regime. The characteristic life-time of the domains in the
movie is $\tau\lesssim 20\mu s$. We performed simulations with
different ratios of translation vs.~state-exchange move attempts (see
simulation details, above) and found that increasing the fraction of
state exchange move attempts has an insignificant impact on the
observed dynamics. This indicates that the association/dissociation of
the clusters is mainly governed by the lateral diffusion of the lipids
rather than by the change in their states.

\begin{figure}
\centering
\includegraphics[width=8.75cm]{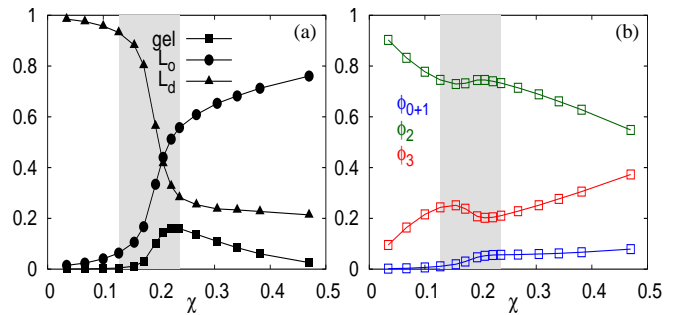}
\caption{ (a) The fractions of sites belonging to the $L_d$, $L_o$,
  and gel regions. (b) The composition of sites in the $L_0$ state:
  $\phi_{0+1}$ is the fraction of sites that are occupied by voids and
  disordered chains, $\phi_2$ by ordered chains, and $\phi_3$ by
  Chol. The gray-shaded area marks the $L_d+L_o$ coexistence
  regime. Both figures are plotted as function of $\chi$ at
  $T=316.5K$.}
\label{fig3}
\end{figure}

Figs.~\ref{fig1}(h1) and (h2) provide a magnified view of the regions
marked by squares in snapshots (e) and (d), respectively. These
regions contain about 500 lipids which are plotted with the same
colors coding as in the larger snapshots. Each point is also marked by
a colored border, indicating whether the site is part of the $L_d$
region (purple), $L_0$ region (yellow), or a gel-like cluster of
hexagonally-packed ordered chains (black) [labeled as ``hexag.'' in
  (h2)]. The algorithm with which the different regions are identified
is described in the Supplemental Material~\cite{supp}.
Fig.~\ref{fig1}(h3) presents a snapshot of similar size from an
atomistic simulation study of DPPC/Chol
mixtures~\cite{javanainen17}. [The labels ``$L_d$'', ``$L_o$'', and
  ``hexag.'' appear in the original copy and were adopted by us in
  snapshot (h2).]  The similarity between the results of the atomistic
(h3) and lattice simulations (h1,h2) is striking. In both, we observe
Chol-free gel-like nano-domains surrounded by regions with high
density of Chol and ordered chains.  The composition of the system,
i.e., fraction of sites belonging to each of the three regions
(averaged over time and over the entire lattice) are plotted in
fig.~\ref{fig3}(a) as a function of $\chi$ at $T=316.5K$. The data
shows that the addition of Chol leads to a monotonic increase
(decrease) in the population of the $L_o$ ($L_d$) states, with the
sharpest incline (decline) occurring in the $L_d+L_o$ coexistence
regime (marked by a gray shaded area). The fraction of sites in the
gel-like clusters exhibits a curious non-monotonic behavior: It
increases from zero as the system enters the $L_d+L_o$ regime and then
drops back to zero as $\chi$ continues to grow.

To conclude, our findings on the inhomogeneity of the $L_o$ domains
are consistent with previous experimental~\cite{amstrong13} and
atomistic simulations~\cite{javanainen17} studies. They suggest that
the gel-like regions should be treated separately and not be counted
as part of the $L_0$ state. The composition of the latter is plotted
in fig.~\ref{fig3}(b). It contains ordered chains ($\phi_2$) and Chol
($\phi_3$), with a small amount of voids and disordered chains
($\phi_{0+1}$).

{\bf Discussion and Summary:} Formation of $L_o$ domains has been
associated with three general mechanisms~\cite{schmid17}: (i)
curvature-composition coupling, (ii) microemulsion stabilization by
lineactant hybrid lipids, and (iii) near-critical fluctuations. None
of these mechanisms seems to be relevant to the system under
investigation.  The model involves no curvature energy term (first
mechanism), shows no evidence for substantial presence of lipids with
ordered and disordered chains along the interfaces between $L_d$ and
$L_o$ regions (second mechanism). With only the first two terms in
Eq.~(\ref{eq:mcenergy}), the system Hamiltonian can be mapped onto the
Ising model. However, we find no signs of criticality (third
mechanism) near the edge of the $L_d+L_o$ coexistence regime [point
  (f) in fig.~\ref{fig1}], suggesting that with the addition of Chol
[third term in Eq.~(\ref{eq:mcenergy})] and the elastic energy term
(fourth term), the model no longer belongs to the Ising universality
class.

Our results call for a new thermodynamic explanation of the formation
of the $L_0$ domains and their internal inhomogeneous structure.
Adding Chol molecules to the membrane in the $L_d$ phase ($T>T_m$)
causes an increasing number of chains in their vicinity to flip from a
disordered to an ordered state, with which Chol interacts more
favorably. However, the attraction between the Chol and the ordered
chain is weaker than the interaction between the ordered chains with
each other ($\epsilon_{23}<\epsilon_{22}$), and provides only partial
compensation for the loss of chain configurational entropy. Therefore,
as the concentration of Chol and ordered chains somewhat grows, the
later partially segregate and form the gel-like clusters. The
distribution of lipids and Chol in fig.~\ref{fig1}(h1)-(h3) may be
interpreted as a coexistence of regions in the gel and $L_d$ phases
which, close to the first-order melting temperature
[$(T-T_m)/T_m\lesssim 0.02$], are nearly-equivalent
thermodynamically. Chol is excluded from the gel and little dissolves
in the $L_d$ region. It accumulates at high concentration along the
gel-$L_d$ interfaces and induces ordering of the lipid chains,
creating a $L_o$ ``buffer'' region. This may be regarded as if the
Chol molecules serve as lineactants between the ordered and disordered
lipid chains.  The composition of $L_o$ state does not change much in
the coexistence region of the phase diagram [see fig.~\ref{fig3}(b)],
suggesting that it may be regarded as a metastable phase. Indeed, a
single chain exchange from an ordered to a disordered state (while
keeping unchanged the states of the others) results in increase in the
system free energy for most of the chains in the $L_o$ phase (see data
in the Supplemental Material~\cite{supp}). This explains the prolonged
lifetime of the domains. At higher values of $\chi$, Chol ``invades''
the gel clusters and eliminates them [fig.~\ref{fig3}(a)], and the
$L_o$ domains merge into a single $L_o$ phase.

To summarize, our model provides a dynamical picture of DPPC/Chol
mixtures over an unprecedented wide range of length- and
time-scales. Its minimal nature helps us identifying the disparity
between the packing interactions of ordered chains with Chol and with
each other, as the reason for their non-ideal mixing within the $L_o$
nano-domains. At larger spatial scales, there is no macroscopic
$L_d/L_o$ phase separation. The system arranges itself such that the
Chol-rich $L_0$ regions separate the gel and Chol-poor $L_d$
regions. The basic thermodynamic mechanism proposed here may also
apply to ternary model mixtures of saturated and unsaturated lipids
with Chol (where, indeed, similar heterogeneous domains have also been
recently identified~\cite{sodt14}) and may be relevant to raft domains
in biological membranes.

\begin{acknowledgments}This work was supported by the Israel Science
Foundation (ISF), grant No.~991/17. TS thanks the Planning and
Budgeting Committee of the Council for Higher Education (Israel) for
supporting his post-doctoral fellowship.
\end{acknowledgments}


\begin{thebibliography}
\medskip


\bibitem{alberts-book} B. Alberts, A. Johnson, J. Lewis, M. Raff,
  K. Roberts, and P. Walter, {\em Molecular Biology of the Cell}, 4th
  edition (Garland Science, New York, 2002).

\bibitem{simons97} K. Simons and E. Ikonen, Functional rafts in cell membranes\/,
 Nature {\bf 387}, 569 (1997).

\bibitem{pike06} L. Pike, Rafts defined: A report on the keystone symposium on lipid rafts and cell function\/, 
J. Lipid Research
  {\bf 47}, 1597 (2006).

\bibitem{levental20} I. Levental, K. R. Levental, F. A. Heberle, Lipid
  rafts: Controversies resolved, mysteries remain\/, Trends Cell
  Biol. {\bf 30}, 341 (2020).
  
\bibitem{simons00} K. Simons and D. Toomre, Lipid rafts and signal transduction\/,
 Nat. Rev. Mol. Cell Biol. {\bf 1}, 31 (2000).

\bibitem{leitinger02} B. Leitinger and N. Hogg, The involvement of lipid rafts in the regulation of integrin function\/, 
J. Cell Sci. {\bf 115}, 963 (2002).  

\bibitem{murai12} T. Murai, The role of lipid rafts in cancer cell adhesion and migration\/,
 Int. J. Cell Biol. {\bf 2012},
  763283 (2012).

\bibitem{ikonen01} E. Ikonen, Roles of lipid rafts in membrane transport\/,
 Curr. Opin. Cell. Biol. {\bf 13}, 470 (2001).

\bibitem{hanzal07} M. F. Hanzal-Bayer and J. F. Hancock, Lipid rafts and membrane traffic\/,
 FEBS Lett. {\bf 581}, 2098 (2007).

\bibitem{vdgoot01} F. G. van der Goot and T. Harder, Raft membrane domains: From a liquid-ordered membrane phase to a site of pathogen attack\/,
 Semin. Immunol. {\bf 13}, 89 (2001).

\bibitem{kaiser09} H.-J. Kaiser, D. Lingwood, I. Levental,
  J. L. Sampaio, L. Kalvodova, L. Rajendran, and K. Simons, Order of lipid phases in model and plasma membranes\/,
  Proc. Natl. Acad. Sci. U.S.A {\bf 106}, 16645 (2009).
  
\bibitem{filippov04} A. Filippov, G. Or\"{a}dd, and G. Lindblom, Lipid lateral diffusion in ordered and disordered phases in raft mixtures\/,
 Biophys. J. {\bf 86}, 891 (2004).

\bibitem{holl08} M. M. B. Holl, {\em Cell plasma membranes and phase
  transitions}\/, in: {\em Phase Transitions in Cell Biology}\/,
  G. H. Pollack and W. C. Chin (eds) (Springer, Dordrecht, 2008).

\bibitem{komura04} S. Komura, H. Shirotori, P. D. Olmsted, and
  D. Andelman, Lateral phase separation in mixtures of lipids and cholesterol\/,
 Europhys. Lett.  {\bf 67}, 321 (2004).

\bibitem{feigenson07} G. W. Feigenson, Phase boundaries and biological membranes\/,
 Annu. Rev. Biophys. Biomol. Struct. {\bf
  36}, 6377 (2007).  

\bibitem{veatch07} S. L. Veatch, O. Soubias, S. L. Keller and
  K. Gawrisch, Critical fluctuations in domain-forming lipid mixtures\/,
 Proc. Natl. Acad. Sci. U.S.A {\bf 105}, 17650 (2007).

\bibitem{goni08} F. M. Go\~{n}i, A. Alonso, L. A. Bagatolli,
  R. E. Brown, D. Marsh, M. Prieto, and J. L. Thewalt, Phase diagrams of lipid mixtures relevant to the study of membrane rafts\/,
 Biochim. Biophys. Acta Mol. Cell Biol. Lipids {\bf
    1781}, 665 (2008).  

 \bibitem{putzel08} G. G. Putzel and M. Schick, Phenomenological model and phase behavior of saturated and unsaturated lipids and cholesterol\/,
 Biophys. J. {\bf 95}, 4756 (2008). 
  
\bibitem{chong09} P. L.-G. Chong, W. Zhu, and B. Venegas, On the lateral structure of model membranes containing cholesterol\/,
  Biochim. Biophys. Acta Biomembr. {\bf 1788}, 2 (2009).

\bibitem{feigenson09} G. W. Feigenson, Phase diagrams and lipid domains in multicomponent lipid bilayer mixtures\/,
  Biochim. Biophys. Acta Biomembr. {\bf 1788}, 47 (2009).
  
\bibitem{schick12} M. Schick, Membrane heterogeneity: Manifestation of a curvature-induced microemulsion\/,
 Phys. Rev. E
  {\bf 85}, 031902 (2012).

\bibitem{komura14} S. Komura and D. Andelman, Physical aspects of heterogeneities in multi-component lipid membranes\/,
 Adv. Colloid
  Interface Sci. {\bf 208}, 34 (2014).

\bibitem{sadeghi14} S. Sadeghi, M. M\"{u}ller, and R. L. C. Vink, Raft formation in lipid bilayers coupled to curvature\/,
    Biophys. J. {\bf 107}, 1591 (2014).
    
\bibitem{sodt14} A. J. Sodt, M. L. Sandar, K. Gawrisch, R. W. Pastor,
  and E. Lyman, The molecular structure of the liquid-ordered phase of lipid bilayers\/,
 J. Am. Chem. Soc. {\bf 136}, 725
  (2014).

\bibitem{ispen87} J. H. Ipsen, G. Karlström, O. G. Mourtisen,
  H. Wennerstr\"{o}m, and M. J. Zuckermann, Phase equilibria in the phosphatidylcholine-cholesterol system\/,
  Biochim. Biophys. Acta Biomembr. {\bf 905}, 162 (1987).

\bibitem{vist90} M. R. Vist and J. H. Davis, Phase equilibria of cholesterol/dipalmitoylphosphatidylcholine mixtures: Deuterium nuclear magnetic resonance and differential scanning calorimetry\/,
  Biochemistry {\bf 29}, 451 (1990).
  
\bibitem{schmid17} F. Schmid, Physical mechanisms of micro- and nanodomain formation in multicomponent lipid membranes\/,
Biochim. Biophys. Acta Biomembr. {\bf 1859}, 509 (2017).

\bibitem{rhein13} M. C. Rheinst\"{a}dter and O. G. Mouritsen, Small-scale structure in fluid cholesterol–lipid bilayers\/,
  Curr. Opin. Colloid Interface Sci. {\bf 18}, 440 (2013).

\bibitem{amstrong13} C. L. Armstrong, D. Marquardt, H. Dies, N.
  Ku\v{c}erka, Z. Yamani, T. A. Harroun, J. Katsaras, A.-C. Shi, amd
  M. C. Rheinst\"{a}dter, The observation of highly ordered domains in membranes with cholesterol\/,
 PLoS ONE {\bf 8}, e66162
  (2013).
    
\bibitem{javanainen17} M. Javanainen, H. Martinez-Seara, and
  I. Vattulainen, Nanoscale membrane domain formation driven by cholesterol\/,
 Sci. Rep. {\bf 7}, 1143 (2017).

\bibitem{pink80} A. Caill\'{e}, D. Pink, F. De Verteuil, and
  M. J. Zuckermann, Theoretical models for quasi-two-dimensional mesomorphic monolayers and membrane bilayers\/,
  Can. J. Phys. {\bf 58}, 581 (1980).

\bibitem{almeida11} P. F. Almeida, A simple thermodynamic model of the liquid-ordered state and the interactions between phospholipids and cholesterol\/,
 Biophys. J. {\bf 100}, 420 (2011).
 

\bibitem{klump81} H. H. Klump, B. P. Gaber, W. L. Peticolas, and P. Yager, Thermodynamic properties of mixtures of deuterated and undeuterated dipalmitoyl phosphatidylcholines (differential scanning calorimetry/lipid bilayers/membranes)\/,
  Thermochim. Acta {\bf 48}, 361 (1981).
 
\bibitem{smond99} A. M. Smondyrev and M. L. Berkowitz, Structure of dipalmitoylphosphatidylcholine/cholesterol bilayer at low and high cholesterol concentrations: Molecular dynamics simulation\/,
  Biophys. J. {\bf 77}, 2075 (1999),


\bibitem{doniach78} S. Doniach, Thermodynamic fluctuations in phospholipid bilayers\/,
  J. Chem. Phys. {\bf 68}, 4912 (1978).

\bibitem{almeida09} P. F. F. Almeida, Thermodynamics of lipid interactions in complex bilayers\/,
    Biochim. Biophys. Acta Biomembr. {\bf 1788}, 72 (2009).
  
\bibitem{almeida2018} P. F. F. Almeida, How to determine lipid interactions in membranes from experiment through the Ising model\/,
 Langmuir {\bf 35}, 21(2018).

 
\bibitem{supp} See Supplemental Material at (URL will be inserted by
  the publisher) for more details about the lattice site classification algorithm, the stability analysis of the ordered lipid chains in the liquid ordered region and a simulation movie showing the dynamics of the
  liquid ordered domains.
  


\end{thebibliography}
\renewcommand{\refname}{{Bibliography}}

\end{document}